# Extended X-ray absorption fine structure study of the Er bonding in AlNO:Er$_x$ films with $x \leq 3.6\%$

M. Katsikini,[1] V. Kachkanov,[2]†, P. Boulet,[3] P. R. Edwards,[4] K. P. O'Donnell,[4]

V. Brien[3,a)]

[1] Department of Physics, Aristotle University of Thessaloniki, GR-54124 Thessaloniki, Greece.

[2] Diamond Light Source Ltd, Diamond House, Chilton, Didcot, Oxfordshire, OX11 0DE, UK.

[3] Institut Jean Lamour, UMR 7198, CNRS, Université de Lorraine, Campus Artem, 2 allée André Guinier, BP 50840, 54011 Nancy, France.

[4] SUPA Department of Physics, University of Strathclyde, 107 Rottenrow, Glasgow, G4 0NG, Scotland, United Kingdom.

[a)] E-mail: valerie.brien@univ-lorraine.fr

†Now at Tokamak Energy Ltd, 120A Olympic Avenue, Milton Park OX14 4SA, UK

**Abstract**

The structural properties of Er-doped AlNO epilayers grown by radio frequency magnetron sputtering were studied by Extended X-ray Absorption Fine Structure (EXAFS) spectra recorded at the Er $L_3$ edge. The analysis revealed that Er substitutes for Al in all the studied samples and the increase in Er concentration from 0.5 to 3.6 at.% is not accompanied by formation of ErN, Er$_2$O$_3$ or Er clusters. Simultaneously recorded X-ray Absorption Near Edge Structure (XANES) spectra verify that the bonding configuration of Er is similar in all studied samples. The Er-N distance is constant at 2.18-2.19 Å i.e. approximately 15% larger than the Al-N bondlength, revealing that the introduction of Er in the cation sublattice causes considerable local distortion. The Debye-Waller factor, which measures the static disorder, of the second nearest shell of Al neighbors, has a local minimum for the sample containing 1% Er that coincides with the highest photoluminescence efficiency of the sample set.

**I. Introduction**

The doping of erbium (Er) in semiconductors has mainly been motivated by the fact that the infra-red emission band of Er ions coincides with the local minimum in the optical absorption spectrum of silica optical fibers. The capacity to integrate nitrides into current or future devices and the reduced temperature quenching of wide band gap semiconductors give Er-doped AlN significant promise for other industrial applications. The photo- and cathodo-luminescence of AlNO:Er$_x$ ($x = 0.5 - 3.6\%$) samples was studied as a function of the Er level in the visible and infra-red domains.[1,2] The cathodoluminescence work led to an interpretation of the optical mechanisms. It appears that the concentration quenching of AlNO:Er is due to different mechanisms than the ones encountered in GaN:Er, confirming previous observations by other authors.[3] Whereas the efficiency of the latter (and the electroluminescent devices to come) seems to be limited by the finite solubility of Er in GaN, as attested by the formation of Er rich nitrides or clusters,[4] AlNO:Er appears to form a solid solution over the quoted range. The observed quenching was interpreted with Förster's theory leading to electrostatic quadrupole-dipole interactions[1] and observations made by Scanning Transmission Electron Microscopy X-ray-mapping imagery, conventional Transmission Electron Microscopy and X-ray diffraction of the samples could not detect clusters or precipitates.[5] In order to deepen our comprehension of the quenching, we present here a study of the local bonding configuration of the Er ion as a function of its concentration in AlNO. X-ray Absorption Fine Structure (XAFS) spectroscopy is an ideal technique for the study of the local bonding environment of rare earth dopants, as it does not necessitate long range order periodicity. Detailed information on the local order is expected to help progress the understanding of the physical optical mechanisms of AlNO:Er samples prepared by reactive Radio Frequency (R.F) magnetron sputtering that were earlier shown to form a solid solution.[5]

## II. Growth conditions and experimental details

The AlNO:Er samples are 500 nm - thick films deposited on silicon substrates by R.F. reactive magnetron sputtering at room temperature. Er doping of the samples was obtained by co-sputtering Er with Al. The resulting samples are polygranular, exhibiting a columnar morphology with an average grain width of 20 – 30 nm measured arbitrarily at a thickness of approximately 300 nm. The shape of the columns is slightly conical, thinner near the substrate and widening as the film was grown. The growth modes of these samples have been extensively studied and reported in Refs. 6,7. The variation of the average Er level from one layer to another was obtained by using an aluminum target on which Er pellets were placed. The progressive consumption and depletion of the target from the dopant ensured that the samples were gradually poorer in the rare earth as successive films were deposited. It should be pointed out that the variation in the Er concentration among the samples did not affect the average width of the columns, as has been verified by transmission electron microscopy cross section images.[5] The impact of the nanogranularity of the polycrystalline films on the EXAFS data is therefore expected to be limited, as grain size effects are generally important only for more distant shells than the first and the second ones and for cluster sizes smaller than 10 – 15 nm.[8] More details on the morphological characterization of the samples studied here can be found elsewhere.[5] Five samples were prepared with oxygen content limited to 5 – 8 at.% with a variable average content of Er ranging from 0.5 to 3.6 at.%. All films were thermally treated post-deposition, employing conditions optimized to "activate" the luminescence (90 min at 950°C) as recommended in the literature.[9,10]

The chemical compositions of the AlNO:Er$_x$ ($x$ = 0.5 – 3.6 %) samples were obtained by Energy Dispersive X-ray (EDX) spectroscopy and by Rutherford Backscattering Spectrometry (RBS). The EDX analysis was performed by means of a Princeton Gamma-Tech (PGT) spectrometer mounted on a CM20 Philips microscope and equipped with an ultra-thin window X-ray detector. The experiments were carried out in nano-probe mode with a diameter of the probe of 10 nm. The Cliff Lorimer coefficients were determined by using the stoichiometric compounds $Er_2O_3$, $Al_2O_3$ and ErSiNi and the nitride $Si_3N_4$ as reference samples. More details on the methodology can be found in Ref. 11. The calibration of the EDX measurements was validated by independent RBS measurements. The compositional and crystallochemical characteristics of the samples are compiled in Table I.

XAFS measurements near the Er $L_3$ edge at 8357.9 eV were performed at the Diamond Light Source B18 beamline at room temperature.[12] X-rays produced by a bending magnet were collimated by a Pt-coated collimating mirror and tuned to the desired energy by a Si(111) double crystal monochromator. The X-ray beam was focused to a spot size of 500 μm (horizontal) by 300 μm (vertical) using a Pt-coated double-bend focusing mirror. A pair of Pt coated mirrors was used for harmonic rejection. The Er $L\alpha_1$ X-ray fluorescence line at 6948.7 eV was monitored by a 9-element monolithic Ge detector with XSPRESS-II readout electronics. The Si(111) monochromator was calibrated using the Cr $K$-edge at 5989 eV prior to the measurements. The incidence angle of the X-ray beam on the sample surface was 45°.

**III. Results and discussion**

The Er $L_3$ edge XANES spectra of the studied samples are shown in Fig. 1. They were subjected to subtraction of the pre-edge linear background and normalization to the edge-jump. They show a prominent characteristic "white line" that is attributed to $2p_{3/2} \rightarrow 5d$ transitions. The position of the "white line" is approximately 8362 eV whereas a second broad feature appears at approximately 30 eV above. The spectra differ considerably from those of $Er_2O_3$ reported by Ishii *et al.*[13] and Mao *et al.*[14], thereby excluding the formation of pure $Er_2O_3$ phases around Er in the studied samples.

Detailed information on the bonding configuration of the Er dopants was obtained from the Er $L_3$ edge EXAFS spectra. Prior to fitting, spectra were subjected to background subtraction and transformation to k-space using the ATHENA software.[15] The resulted $\chi(k)$ spectra were fitted, in both the k- and R-space, using the FEFFIT software with scattering paths that were constructed with the FEFF8 code.[16] The selection of the fitting model was based on previous investigations on the samples. Indeed, XRD data[5] revealed that the samples under study are solid solutions of Er in aluminum oxynitride, AlNO:Er$_x$ ($x \leq 3.6$ %), characterized by an invariant crystallographic space group and a linear dependence of their lattice constants on the Er concentration, as shown in Fig. 2. The analysis of the XRD intensities as a function of the Er content led to the deduction that the Er ions substitute mainly for Al. However, partial occupation of octahedral and/or tetrahedral insertion sites of the wurtzite structures by the Er atoms was also detected.[5] Furthermore, calculations performed using Density Functional Theory (DFT) predicted that the rare earth atoms are distributed throughout the lattice and do not agglomerate, and confirmed the main occupation of Er in the Al sites, with much lower occupancies of insertion sites (octahedral and to a lesser extent tetrahedral).[17] Therefore, the crystallographic model that was used in the analysis of the EXAFS spectra was wurtzite AlN (space group P6$_3$mc), where Er substitutes for Al. It should be pointed out that in addition to this substitutional model, the following models were also considered: Er metal,[18] Er$_2$O$_3$,[19] ErN,[20] Er occupying octahedral voids located in 2a Wyckoff positions (0, 0, $z$) or tetrahedral voids located in 2b Wyckoff positions (1/3, 2/3, $z$) of the wurtzite AlN structure. For Er located in wurtzite structural voids and in Al substitutional sites, the relaxed positions, obtained by DFT calculations published in a comprehensive work by our colleagues, were taken into account.[17] The Fourier transform amplitudes of the theoretical k$^3$-weighted $\chi$(k) spectra, obtained using the above mentioned models, are compared with the experimental spectrum of the AlNO film that contains 2.8% Er in Fig. 3. The theoretical spectra were calculated using the FEFF8 code and damping of the EXAFS signal due to thermal disorder was included in the calculations by exploiting the Debye model. The Debye temperatures reported in the literature for Er metal, Er$_2$O$_3$ and ErN (163, 380 and 210 K, respectively) were used.[21,22,23] For the point defects (Er that substitutes for Al or located in voids) the Debye temperature of AlN (835 K) was used.[24] Although the use of the Debye model works well for crystalline semiconductors,[25] it is not expected to be very accurate for the case of point defects. Therefore, only the positions of Fourier transform maxima can be used for comparison with the experimental data. It should be mentioned that the peak positions are not corrected for phase shifts, in either the theoretical or experimental spectra.[26] It is evident from Fig. 3 that the case of Er substituting for Al is more consistent with the experimental spectrum compared with the other cases. Furthermore, there is no evidence for Er-Er bonding at the expected Er-Er inter-atomic distance of 3.47–3.56 Å.

The $\chi(k)$ spectra and the corresponding Fourier transform amplitudes of the studied samples are shown in Fig. 4. The fitting results are listed in Table II and verify that Er is incorporated in the AlN lattice substituting for Al. The dependences of the nearest neighbor distances and Debye-Waller factors on the Er content are shown in Fig. 5. In Figs. 5(a,b) the Er-N and Er-Al inter-atomic distances are compared with the corresponding values obtained using the $a$ and $c$ lattice constants, as determined by XRD, and assuming the validity of the virtual crystal approximation (VCA). According to VCA, all atoms in an alloy occupy ideal lattice sites obtained using the lattice constants.[27] The results suggest that substitution of Al by Er results in considerable local distortion due to the large difference between the radii of Er and Al. More specifically, the Er-N distance, in the range 2.18 to 2.19 Å, is approximately 15% larger than the Al-N distance determined using the XRD lattice constants. On the other hand, the Er-Al distance (second nearest neighboring shell) is found ≈2% larger than the Al-Al distance, revealing that Er causes local distortion that is mainly accommodated by bond bending. Similar behavior has been reported in the literature in the case of Er doped GaN films with Er content lower than 0.4%.[4,28] In these studies, the Er-N distance was found equal to 2.14 to 2.23 Å, revealing that the Er-N distance is mainly determined, in both AlN and GaN, by the size of the Er atom and molecular orbital formation constraints. The reported Er-Ga distance in the Er-doped GaN was 3.23–3.26 Å i.e. 2% larger than the Ga-Ga distance in undoped GaN. Higher Er contents resulted in the formation of ErN inclusions.[4] It should be also pointed out that the Er-N distance in AlN determined in this study is by ≈10 % smaller than the Er-N distance in ErN.[28] However, in the latter case Er is octahedrally coordinated with 6 N atoms, contrasting with the tetrahedral coordination of Er in AlN. Substitution of Al by Er atoms in Er implanted AlN has been also demonstrated by EXAFS spectroscopy at the Er $L_3$ edge. An average number of 3-4 N neighboring atoms of Er, found at the distance of 0.22 nm, has been reported.[29] The dependence of the Debye-Waller factors that account for the static and thermal disorder on the Er content [Fig. 5(c)] exhibits a local minimum for the Er-Al atom pair in the case of the sample that contains 1% Er. The lower static disorder of this sample is probably correlated to the higher photoluminescence efficiency observed for AlN samples doped with 1% Er,[2] revealing that the homogeneous bonding configuration of Er in AlN might play an important role in the light emission properties.

Many compounds doped with rare earths (RE) are subject to RE nanoprecipitation partly responsible for concentration quenching. Whereas previous experimental investigations and DFT calculations on our system[17] suggests that no precipitation occurs, the EXAFS study presented here allows us to draw stronger conclusions on that point. No Er-rich clusters such as ErN or $Er_2O_3$ were detected by XAFS. Although it is a difficult issue, the minimum detection limit for secondary phases by EXAFS here can be roughly estimated below 10 % at. Theoretical EXAFS spectra were constructed as weighted average of the simulated spectra shown in Fig. 3, of Er that substitutes for Al in AlN and Er located in a secondary phase or defect configuration ($Er_2O_3$, ErN, metallic Er and Er in octahedral or tetrahedral insertion site of AlNO wurtzite). Fitting quality factors (r-factors) of these theoretical spectra can be found in the supplementary material. In all cases 10% of the secondary crystallographic configuration causes a worsening of the r-factor by more than 20%, and unreasonable coordination numbers and Debye-Waller factors were obtained. The local order and the decoration of the surrounding shells of the Er atoms probed here confirm the presence of a AlNO:Er solid solution. They show that the magnetron sputtered AlNO:$Er_x$ ($x = 0.5 - 3.6\%$) films do contain phases that were predicted to be thermodynamically stable by DFT calculations even when Er concentration reaches nearly 4 at. %.[17] These results also imply that XRD detexturation results[5] obtained on the same thin films need re-examination: the exploitation of relative intensities of the peaks to conclude about the relative distribution of Er in insertion sites of wurtzite has to be reconsidered. It then appears that, as already encountered in other thin films adopting the hexagonal ZnO structure such as $ZnSnN_2$[30] or $ZnGeN_2$[31], the simple hexagonal unit cell is not sufficient. Even the ortho-hexagonal distortion often used[32] in the case of the ternary zinc-tin nitride is not satisfactory in most of the cases. For AlN doped systems, more advanced structure analyses, using single crystal XRD or precise XRD powder data, would be necessary to determine the correct structural description.

**IV. Conclusions**

A series of Er- ($x < 3.6\%$) doped AlNO samples, with oxygen content ranging from 5 to 8 at. %, were studied by means of XAFS spectroscopy at the Er $L_3$ edge. The analysis results reveal that Er predominantly substitutes for Al in the wurtzite structure having tetrahedral coordination, and that AlNO:Er$_x$ samples are solid solutions over the whole studied composition range. The incorporation of Er in the host lattice causes considerable distortion as the Er-N distance is found to be 15% longer than the Al-N bond length. The local stress thus induced is mainly accommodated by bond bending, as evidenced by the second nearest neighbor distances, and could be the origin of the unresolved structural distortion. Finally, the Debye-Waller factor in the second nearest neighboring shell, corresponding to the Er-Al atomic pair, exhibits a local minimum at $x$ = 1% and might be correlated to the better luminescence efficiency of this sample.

**Supplementary material:** Video related to the minimum detection limit of the secondary phases.

**Acknowledgements**

Diamond Light Source is acknowledged for providing beamtime via proposal SP9370-1. J. Ghanbaja is thanked for performing the EDX observations. The authors are also grateful to J. Kioseoglou and T. Pavloudis for providing the atomic positions of the defects included in Fig. 3. S.S. Hussain, P. Pigeat are thanked for their contribution with the preparation of the samples. We also acknowledge financial support from the European Union ERASMUS+ program for subsidizing exchanges between French and Greek partners.

**References**

[1] V. Brien, P.R. Edwards, P. Boulet, and K. P. O'Donnell, submitted to Journal of Luminescence (2018).

[2] H. Rinnert, S .S. Hussain, V. Brien, J. Legrand, and P. Pigeat, J. Lumin., **132**, 2367 (2012).

[3] F. Benz, A. Gonser, R. Völker, T. Walther, J.-T. Mosebach, B. Schwanda, N. Mayer, G. Richter, and H. P. Strunk, J. Lumin., **145**, 855 (2014).

[4] V. Katchkanov, J. F. W. Mosselmans, K. P. O'Donnell, E. Nogales, S. Hernandez, R. W. Martin, A. Steckl, and D. S. Lee, Opt. Mater., **28**, 785 (2006).

[5] V. Brien and P. Boulet, Acta Mater., **90**, 37 (2015).

[6] V. Brien and P. Pigeat, J. Cryst. Growth., **299**, 189 (2007).

[7] V. Brien and P. Pigeat, J. Cryst. Growth., **310**, 3890 (2008).

[8] M. Katsikini and E. C. Paloura, XAFS for characterization of nanomaterials. In: C. Kumar (eds.) X-ray and neutron techniques for nanomaterials characterization. pp. 157-246, Springer, Berlin, Heidelberg (2016).

[9] M. Maqbool, Eur. Phys. J. Appl. Phys., **34**, 31 (2006).

[10] J. A. Guerra, F. Benz, A. R. Zanatta, H. P. Strunk, A. Winnacker, and R. Weingärtner, Phys. Status Solidi C, **10**, 68 (2013).

[11] V. Brien, P. Miska, H. Rinnert, D. Geneve, and P. Pigeat, Mater. Sci. Eng. B, **146**, 200 (2008).

[12] A. J. Dent, G. Cibin, S. Ramos, A. D. Smith, S. M. Scott, L. Varandas, M. R. Pearson, N. A. Krumpa, C. P. Jones, and P E Robbins, J. Phys.: Conf. Ser., **190**, 012039 (2009).

[13] M. Ishii, S. Komuro, T. Morikawa, Y. Aoyagi, T. Ishikawa, and T Ueki, Jpn. J. Appl. Phys., **38**, S38-1, 191 (1999).

[14] Y. Mao, J. Bargar, M. Toney, and J. P. Chang, J. Appl. Phys., **103**, 094316 (2008).

[15] B. Ravel and M. Newville, J. Synchrotron Radiat., **12**, 537 (2005).

[16] A. L. Ankudinov, B. Ravel, J. J. Rehr, and S. D. Conradson, Phys. Rev. B, **58**, 7565 (1998).

[17] T. Pavloudis, V. Brien, and J. Kioseoglou, Comput. Mater. Sci., **138**, 128 (2017).

[18] F. H. Spedding, A. H. Daane, and K. W. Herrmann, Acta Crystallogr., **9**, 559 (1956).

[19] A. Saiki, N. Ishizawa, N. Mizutani, and M. Kato, J. Ceram. Soc. Japan, **93**, 649 (1985).

[20] J. C. Fitzmaurice, A. Hector, A. T. Rowley, and I. P. Parkin, Polyhedron, **13**, 235 (1994).

[21] R. E. Skochdopole, M. Griffel, and F. H. Spedding, J. Chem. Phys., **23**, 2258 (1955).

[22] W. R. Manning, O. Hunter, B. R. Powel, J. Am. Ceram. Society, **52**, 436 (1969).

[23] V. Bhalla, D. Devraj, Indian J. Pure Appl. Phys., **54**, 40 (2016).

[24] B. P. Pandey, V. Kumar, and E. Menendez Proupin, Pramana J. Phys., **83**, 413 (2014).

[25] M. Katsikini, F. Pinakidou, E. C. Paloura, P. Komninou, A. Georgakilas, and E. Welter, phys. stat. sol. (a), **205**, 2611 (2008).

[26] P. A. Lee, P. H. Citrin, P. Eisenberger, and B. M. Kincaid, Rev. Mod. Phys. **53**, 769 (1981).

[27] C. S. Schnohr, Appl. Phys. Rev., **2**, 031304 (2015).

[28] P. H. Citrin, P. A. Northrup, R. Birkhahn, and A. J. Steckl, Appl. Phys. Lett. **76**, 2865 (2000).

[29] A. Traverse, Hyperfine Interact. **110**, 159 (1997).

[30] F. Alnjiman, S. Diliberto, J. Ghanbaja, E. Haye, S. Kassavetis, P. Patsalas, C. Gendarme, S. Bruyère, F. Cleymand, P. Miska, P. Boulet, J. F. Pierson, Sol. Energy Mater Sol. Cells, **182**, 30 (2018).

[31] A. M. Shing, Y. Tolstova, N. S. Lewis H. A. Atwater, Sci. Rep., **7**, 11990 (2017).

[32] L. Lahourcade, N. C. Coronel, K. T. Delaney, S.K. Shukla, N.A. Spaldin, H.A. Awater, Adv. Mater., **25**, 2562 (2015).

**Table I:** Compositional and structural characteristics of AlNO:Er$_x$ sputtered films after thermal treatment at 950°C for 90 min. Chemical analyses were performed by EDX. Calibration was done by using discrete compounds, confirmed by RBS on some samples.[2]

| Sample ID | Thickness (nm) | $a$‡ (Å) | $c$‡ (Å) | $c/a$ | Al at. % | N at. % | O at. % | Er at. % |
|---|---|---|---|---|---|---|---|---|
| 1 | 500 | 3.120 | 4.990 | 1.5994 | 42.5 | 52.0 | 5.0 | 0.5* |
| 2 | 500 | 3.130 | 5.000 | 1.5974 | 41.0 | 49.9 | 8.0 | 1.0 |
| 3 | 500 | 3.140 | 5.000 | 1.5924 | 43.0 | 49.0 | 6.6 | 1.4* |
| 4 | 500 | 3.154 | 5.016 | 1.5904 | 41.4 | 49.8 | 6.0 | 2.8 |
| 5 | 500 | 3.162 | 5.207 | 1.5898 | 40.2 | 51.1 | 5.1 | 3.6 |

‡ Obtained by X-ray diffraction (estimated standard deviation: ± 0.005 Å).[5]

* Obtained by RBS analysis.

**Table II:** Fitting results of the EXAFS spectra. $N_i$, $R_i$, and $\sigma^2_i$ stand for the coordination numbers, the nearest neighbor distances and the Debye-Waller factors. i=1 and 2 refer to the first and second nearest neighboring shells that consist of N and Al atoms, respectively. The errors correspond to the fitting uncertainties provided by FEFFIT. The amplitude reduction factor ($S_0^2$) is set equal to 0.73 whereas the energy shift ($\Delta E_0$) was commonly iterated for all the samples and was found equal to -1.8±0.3 eV.

| Er% | $N_1$ | $N_2$ | $R_1$(Å) | $R_2$(Å) | $\sigma^2_1 \times 10^{-3}$ (Å$^2$) | $\sigma^2_2 \times 10^{-3}$ (Å$^2$) |
|---|---|---|---|---|---|---|
| 0.5 | 4.0±0.5 | 8.4±0.9 | 2.18±0.01 | 3.17±0.01 | 7.0±1.5 | 9.4±1.3 |
| 1.0 | 3.7±0.4 | 8.3±0.7 | 2.18±0.01 | 3.16±0.01 | 4.7±1.4 | 5.7±0.9 |
| 1.4 | 4.0±0.8 | 9.1±0.8 | 2.19±0.01 | 3.17±0.01 | 5.3±1.3 | 8.8±1.1 |
| 2.8 | 4.1±0.3 | 9.4±0.7 | 2.19±0.01 | 3.18±0.01 | 3.9±0.8 | 10.3±0.9 |
| 3.6 | 4.1±0.3 | 9.5±0.8 | 2.19±0.01 | 3.18±0.01 | 4.8±0.8 | 12.2±1.1 |

**FIGURE CAPTIONS**

**Figure 1.** Er $L_3$ edge XANES spectra. The position of the "white line" and a broad feature are denoted by vertical lines. The Er content, $x$, is also indicated.

**Figure 2.** $a$ and $c$ lattice constants as determined by XRD.

**Figure 3.** Fourier transform amplitudes of the $k^3$ weighted $\chi(k)$ theoretical spectra of Er, $Er_2O_3$, ErN, interstitial Er in octahedral void [$Er_i$ (oct, valence)], interstitial Er in tetrahedral void [$Er_i$ (tet, valence)], substitutional Er in Al position. The corresponding Fourier transform amplitude for the experimental $\chi(k)$ spectrum of the AlNO sample with 2.8% Er is shown in thick solid line.

**Figure 4:** Fitting of the EXAFS spectra in the (a) k- and (b) R- space. The experimental and fitting curves are shown in thin and thick lines, respectively. The Er content is also indicated.

**Figure 5:** Dependence of the (a) Er-N and (b) Er-Al inter-atomic distances in the Er content, plotted in filled squares and circles, respectively. Empty triangles and lines correspond to the corresponding distances expected by the virtual crystal approximation. (c) Dependence of the Debye-Waller factors of the first and second nearest neighboring shells, shown in empty and filled squares, respectively, on the Er content.

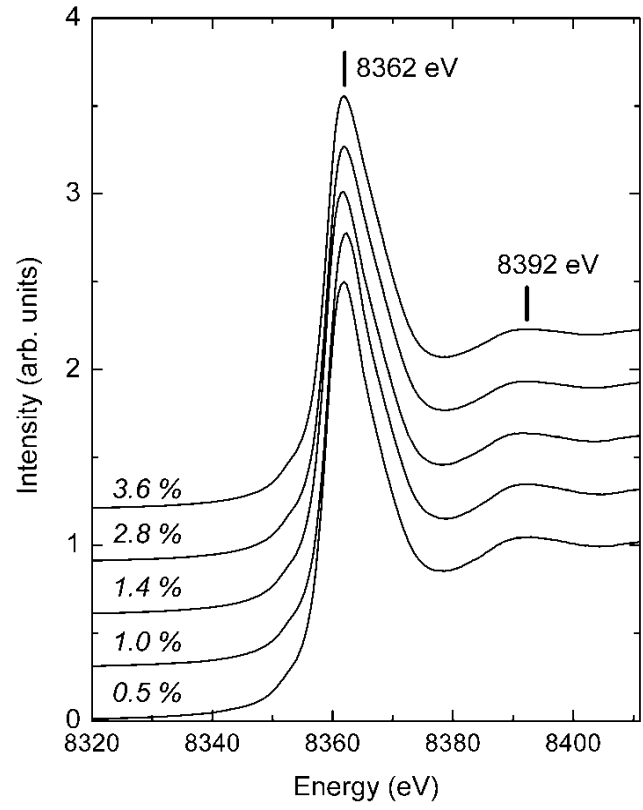

FIGURE 1

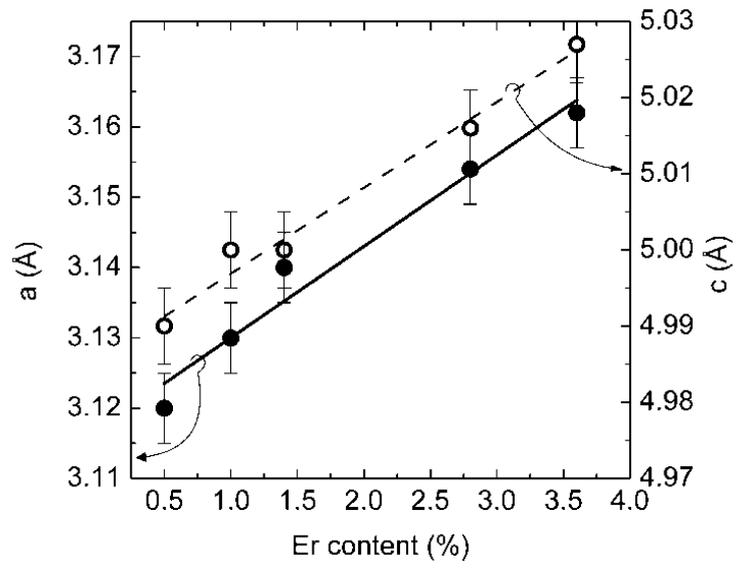

FIGURE 2

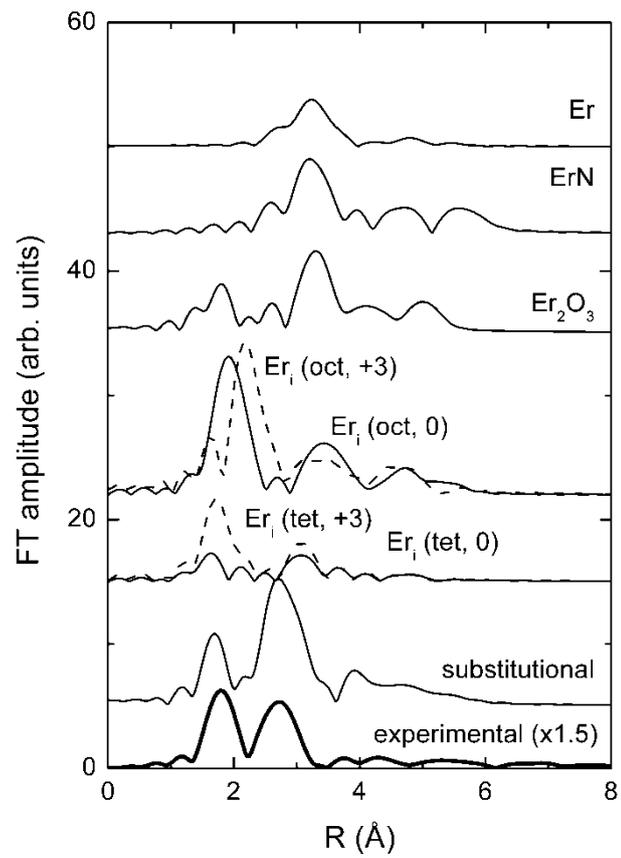

FIGURE 3

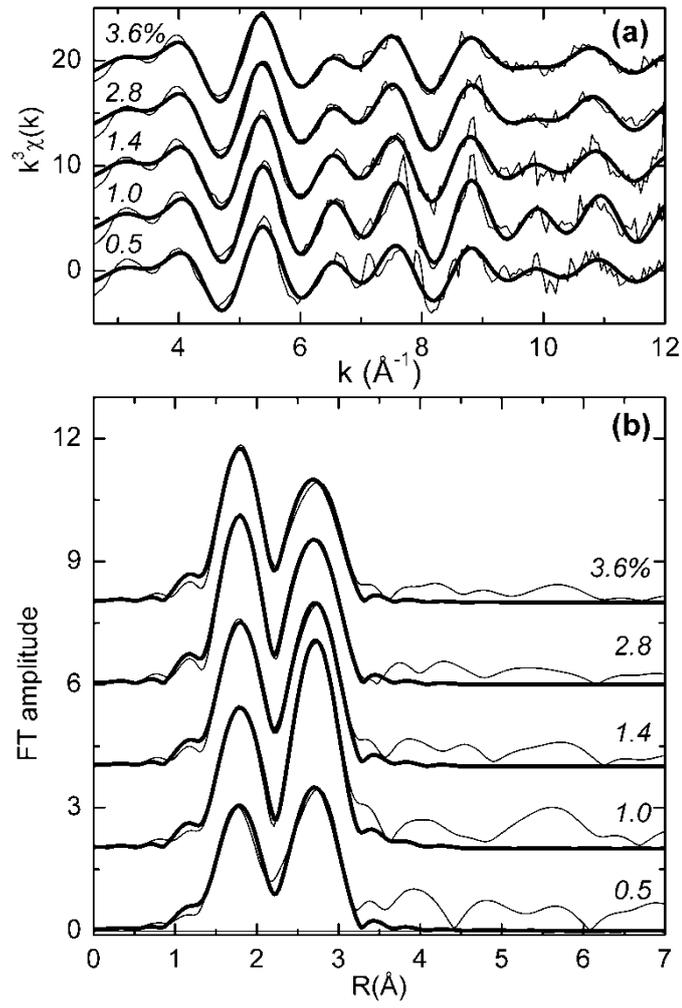

FIGURE 4

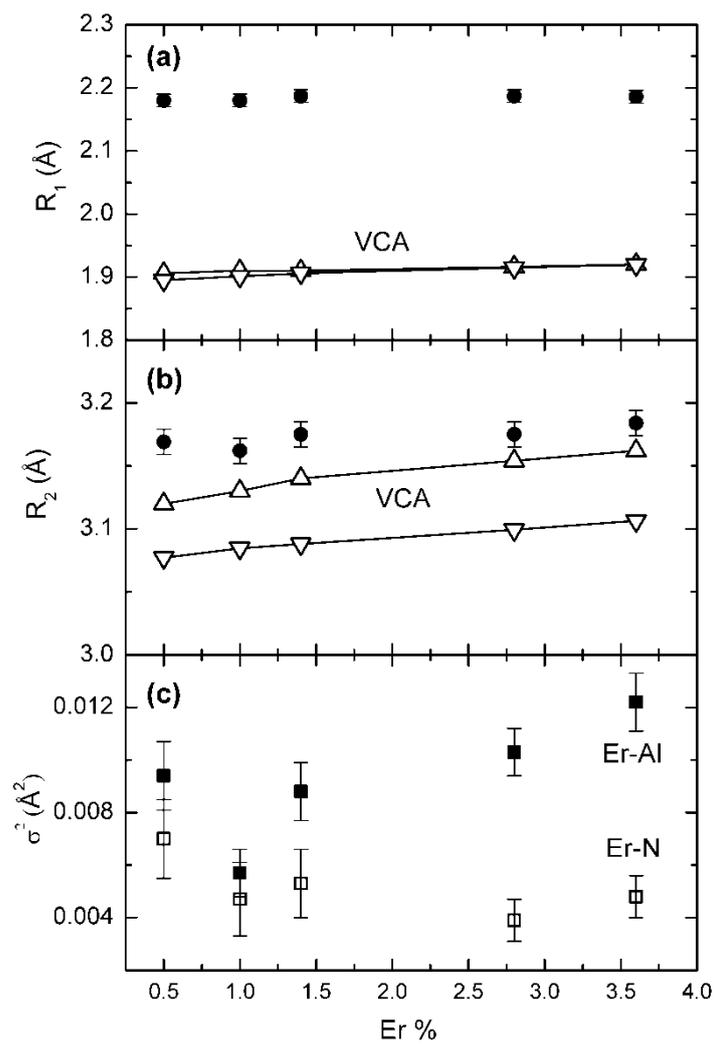

FIGURE 5